\documentclass[preprint,12pt]{imsart}

\usepackage{tikz}
\usepackage{ascmac}
\usepackage{amsmath}
\usepackage{amsfonts}
\usepackage{amsthm}
\usepackage{graphicx}

\usepackage{natbib}
\def\cite{\citet}

\usepackage[dvips]{geometry}
\startlocaldefs

\newtheorem{theorem}{Theorem}
\newtheorem{lemma}[theorem]{Lemma}
\endlocaldefs

\begin{document}

\begin{frontmatter}
\title{An information criterion for auxiliary variable selection in incomplete data analysis}
\runtitle{An information criterion for auxiliary variable selection}
\author[add1,add3]{\fnms{Shinpei} \snm{Imori}\corref{}\ead[label=e1]{imori@hiroshima-u.ac.jp}\thanksref{t1}}
\and
\author[add2,add3]{\fnms{Hidetoshi} \snm{Shimodaira}\corref{}\ead[label=e2]{shimo@i.kyoto-u.ac.jp}\thanksref{t2}} 
\thankstext{t1}{The research was supported in part by JSPS KAKENHI Grant (17K12650) and ``Funds for the Development of Human Resources in Science and Technology'' under MEXT, through the ``Home for Innovative Researchers and Academic Knowledge Users (HIRAKU)'' consortium.}
\thankstext{t2}{The research was supported in part by JSPS KAKENHI Grant (16H02789).}
\address[add1]{Graduate School of Science, Hiroshima University\\
1-3-1 Kagamiyama, Higashi-Hiroshima, Hiroshima 739-8526, Japan\\
\printead{e1}}
\address[add2]{Graduate School of Informatics, Kyoto University\\
Yoshida Honmachi, Sakyo-ku, Kyoto, 606-8501, Japan\\
\printead{e2}}
\address[add3]{RIKEN Center for Advanced Intelligence Project\\
1-4-1 Nihonbashi, Chuo-ku, Tokyo, 103-0027, Japan}

\runauthor{S.~IMORI AND H.~SHIMODAIRA}

\begin{abstract}

Statistical inference is considered for variables of interest, called primary variables, when auxiliary variables are observed along with the primary variables. 
We consider the setting of incomplete data analysis, where some primary variables are not observed. 
Utilizing a parametric model of joint distribution of primary and auxiliary variables, it is possible to improve the estimation of parametric model for the primary variables when the auxiliary variables are closely related to the primary variables. However, the estimation accuracy reduces when the auxiliary variables are irrelevant to the primary variables.
For selecting useful auxiliary variables, we formulate the problem as model selection,
and propose an information criterion for predicting primary variables by leveraging auxiliary variables. The proposed information criterion is an asymptotically unbiased estimator of the Kullback-Leibler divergence for complete data of primary variables under some reasonable conditions. We also clarify an asymptotic equivalence between the proposed information criterion and a variant of leave-one-out cross validation. Performance of our method is demonstrated via a simulation study and a real data example. 

\end{abstract}


\end{frontmatter} 
\section{Introduction}\label{Int}

Auxiliary variables are often observed along with primary variables.
Here, the primary variables are random variables of interest, and our purpose is to estimate their predictive distribution, i.e., a probability distribution of the primary variables in future test data, while the auxiliary variables are random variables that are observed in training data but not included in the primary variables. We assume that the auxiliary variables are not observed in the test data, or we do not use them even if they are observed in the test data. When the auxiliary variables have a close relation with the primary variables, we expect to improve the accuracy of predictive distribution of the primary variables by considering a joint modeling of the primary and auxiliary variables. 

The notion of auxiliary variables has been considered in statistics and machine learning literature. For example, the ``curds and whey'' method \citep{BF97} and the ``coaching variables'' method \citep{TH98} are based on a similar idea for improving prediction accuracy of primary variables by using auxiliary variables. In multitask learning, \citet{C97} improved generalization accuracy of a main task by exploiting extra tasks.
Auxiliary variables are also considered in incomplete data analysis, i.e., a part of primary variables are not observed; \citet{MLM15} showed some theoretical results to make parameter estimation better by utilizing auxiliary variables in Gaussian mixture model (GMM). 

Although auxiliary variables are expected to be useful for modeling primary variables, they can actually be harmful.
As mentioned in \citet{MLM15}, using auxiliary variables may affect modeling result adversely because the number of parameters to be estimated increases and a candidate model of the auxiliary variables can be misspecified. Hence, it is important to select useful auxiliary variables. 
This is formulated as model selection by considering parametric models with auxiliary variables.
In this paper, usefulness of auxiliary variables for estimating predictive distribution of primary variables is measured by a risk function based on the Kullback-Leibler (KL) divergence \citep{KL51} that is often used for model selection.
Because the KL risk function includes unknown parameters, we have to estimate it in actual use. Akaike Information Criterion (AIC) proposed by \citet{A74} is one of the most famous criteria, which is known as an asymptotically unbiased estimator of the KL risk function. AIC is a good criterion from the perspective of prediction due to the asymptotic efficiency; see \citet{S81, S83}. \citet{T76} proposed a modified version of AIC, called Takeuchi Information Criterion (TIC), which relaxes an assumption for deriving AIC, that is, correct specification of candidate model. However, AIC and TIC are derived for primary variables without considering auxiliary variables in the setting of complete data analysis, and therefore, they are not suitable for auxiliary variable selection nor incomplete data analysis.

Incomplete data analysis is widely used in a broad range of statistical problems by regarding a part of primary variables as latent variables that are not observed.
This setting also includes complete data analysis as a special case, where all the primary variables are observed.
Information criteria for incomplete data analysis have been proposed in previous studies. \citet{S94} developed an information criterion based on the KL divergence for complete data when the data are only partially observed.  \citet{CS98} modified the first term of the information criterion of \citet{S94} by the objective function of the EM algorithm \citep{DLR77}. Recently, \citet{SM18} proposed an information criterion, which is derived by mitigating a condition assumed in \citet{S94} and \citet{CS98}. 

However, any of these previously proposed criteria are not derived by taking auxiliary variables into account. Thus, we propose a new information criterion by considering not only primary variable but also auxiliary variables in the setting of incomplete data analysis. The proposed criterion is a generalization of AIC, TIC and the criterion of \citet{SM18}. 
To the best of our knowledge, this is the first attempt to derive an information criterion by considering auxiliary variables. Moreover, we show an asymptotic equivalence between the proposed criterion and a variant of leave-one-out cross validation (LOOCV); this result is a generalization of the relationship between TIC and LOOCV \citep{S77}. 

Note that ``auxiliary variables'' may also be used in other contexts in literature.
For example, \citet{ILH01} considered to use auxiliary variables in missing data analysis,
which is similar to our usage in the sense that auxiliary variables are highly correlated with missing data.
However, they use the auxiliary variables in order to avoid specifying a missing data mechanism;
this goal is different from ours, because no missing data mechanism is considered in our study.  

The reminder of this paper is organized as follows.
Notations as well as a setting of this paper are introduced in Section \ref{pre}.
Illustrative examples of useful and useless auxiliary variables are given in Section \ref{ie}.
The information criterion for selecting useful auxiliary variables in incomplete data analysis is derived in Section \ref{ic}, 
and the asymptotic equivalence between the proposed criterion and a variant of LOOCV is shown in Section \ref{cv}.
Performance of our method is examined via a simulation study and a real data analysis in Sections \ref{simu} and \ref{sec:rd}, respectively.
Finally, we conclude this paper in Section \ref{conc}.
All proofs are shown in Appendix. 

\section{Preliminaries}\label{pre}
\subsection{Incomplete data analysis for primary variables}

First we explain a setting of incomplete data analysis for primary variables in accordance with \citet{SM18}. 
Let $X$ denote a vector of primary variables, which consists of two parts as $X=(Y, Z)$,
where $Y$ denotes the observed part and $Z$ denotes the unobserved latent part.
This setting reduces to complete data analysis of $X=Y$ when $Z$ is empty.
We write the true density function of $X$ as $q_x(x)=q_x(y, z)$ and
a candidate parametric model of the true density as $p_x(x; \theta)=p_x(y, z; \theta)$, where $\theta \in \Theta \subset \mathbb{R}^d$ is an unknown parameter vector and $\Theta$ is its parameter space.
We assume that $x=(y, z) \in \mathcal{Y} \times \mathcal{Z}$ for all density functions,
where $\mathcal{Y}$ and $\mathcal{Z}$ are domains of $Y$ and $Z$, respectively.
Thus the marginal densities of the observed part $Y$ are obtained by
$q_y(y)=\int q_x(y, z) dz$ and $p_y(y; \theta)=\int p_x(y, z; \theta) dz$. 
For denoting densities, we will omit random variables such as $q_y$ and $p_y(\theta)$.
We assume that $\theta$ is identifiable with respect to $p_y(\theta)$. 

In this paper, we consider only a simple setting of i.i.d.~random variables of sample size $n$.
Let $x_i=(y_i, z_i)$, $i=1,\ldots, n$, be independent realizations of $X$,
where we only observe $y_1,\ldots, y_n$ and we cannot see the values of $z_1,\ldots, z_n$.
We estimate $\theta$ from the observed training data $y_1,\ldots, y_n$.
Then the maximum likelihood estimate (MLE) of $\theta$ is given by
\begin{align}
\hat{\theta}_y&={\rm arg}\max_{\theta \in \Theta} \ell_y(\theta) \equiv {\rm arg}\max_{\theta \in \Theta} \frac{1}{n}\sum_{i=1}^n \log{p_y}(y_i; \theta),
\label{est-y}
\end{align}
where $ \ell_y(\theta) $ denotes the log-likelihood function (divided by $n$) of $\theta$ with respect to $y_1, \ldots, y_n$.

If we were only interested in $Y$, we would consider the plug-in predictive distribution $p_y(\hat{\theta}_y)$ by substituting $\hat{\theta}_y$ into $p_y(\theta)$. 
However, we are interested in the whole primary variable $X=(Y,Z)$ and its density $q_x$.
We thus consider $p_x(\hat{\theta}_y)$ by substituting $\hat{\theta}_y$ into $p_x(\theta)$, and evaluate the MLE by comparing $p_x(\hat{\theta}_y)$ with $q_x$.
For this purpose, \citet{SM18} derived an information criterion as an asymptotically unbiased estimator of the KL risk function which measures how well $p_x(\hat{\theta}_y)$ approximates $q_x$.

\subsection{Statistical analysis with auxiliary variables}

Next, we extend the setting to incomplete data analysis with auxiliary variables.
Let $A$ denote a vector of auxiliary variables.
In addition to $Y$, we observe $A$ in the training data, but we are \emph{not} interested in $A$.
For convenience, we introduce a vector of observable variables $B=(Y, A)$ and a vector of all variables $C=(Y, Z, A)$ as summarized in Table~\ref{notation}.
Now $c_i=(y_i, z_i, a_i)$, $i=1,\ldots, n$, are independent realizations of $C$, and
we estimate $\theta$ from the observed training data $b_i=(y_i, a_i)$, $i=1,\ldots, n$.
Let $\hat\theta_b$ be the MLE of $\theta$ by using $A$ in addition to $Y$.
Since we are only interested in the primary variables,
we consider the plug-in predictive distribution $p_x(\hat{\theta}_b)$ by substituting $\hat{\theta}_b$ into $p_x(\theta)$, and evaluate the MLE by comparing $p_x(\hat{\theta}_b)$ with $q_x$.

\begin{table}[htbp]
\centering
\caption{Random variables in incomplete data analysis with auxiliary variables.
$B=(Y,A)$ is used for estimation of unknown parameters, and $X=(Y,Z)$ is used for evaluation of candidate models. }
\begin{tabular}{c|cc|c}
       & Observed & Latent &Complete \\\hline
Primary &$Y$&$Z$&$X$\\
Auxiliary &$A$&--&--\\\hline
All   &$B$&--&$C$
\end{tabular}
\label{notation}
\end{table}

In order to define the MLE $\hat\theta_b$, let us clarify a candidate parametric model with auxiliary variables.
We write the true density function of $C$ as $q_c(c)=q_c(y, z, a)$ and a candidate parametric model of the true density as $p_c(c; \beta)=p_c(y, z, a; \beta)$, where $\beta = (\theta^\top, \varphi^\top)^\top \in \mathcal{B} \subset \mathbb{R}^{d+f}$ is an unknown parameter vector with nuisance parameter $\varphi \in \mathbb{R}^f$
and $\mathcal{B}$ is its parameter space. 
We assume that $c=(y, z, a) \in \mathcal{Y} \times \mathcal{Z} \times \mathcal{A}$ for all density functions,
where $\mathcal{A}$ is the domain of $A$.
We also assume that $\beta$ is identifiable with respect to $p_b(y, a; \beta)=\int p_c(y, z, a; \beta) dz$.
Let us redefine $p_x(\theta)$ as $p_x(y,z ; \theta) = \int p_c(y,z,a ; \beta) da$ and the parameter space of $\theta$ as
\begin{align*}
\Theta = \left\{ \theta ~\middle| \begin{pmatrix}\theta\\ \varphi\end{pmatrix} \in \mathcal{B}\right\}. 
\end{align*}
Then, $\hat\theta_b$ is obtained from the MLE of $\beta$ given by
\begin{align}
\hat{\beta}_b&=\begin{pmatrix}
\hat{\theta}_b\\
\hat{\varphi}_b
\end{pmatrix}={\rm arg}\max_{\beta \in \mathcal{B}} \ell_b(\beta) \equiv {\rm arg}\max_{\beta \in \mathcal{B}} \frac{1}{n}\sum_{i=1}^n \log{p_b}(b_i; \beta),
\label{est-b}
\end{align}
where $ \ell_b(\beta) $ denotes the log-likelihood function (divided by $n$) of $\beta$ with respect to $b_1, \ldots, b_n$.

Finally, we introduce a general notation for density functions.
For a random variable, say $R$, we write the true density function as $q_r(r)$ and a candidate parametric model of $q_r$ as $p_r(r; \theta)$ or $p_r(r; \beta)$.
For random variables $R$ and $S$, we write the true conditional density function of $R$ given $S=s$  as $q_{r|s}(r|s)$ and its corresponding model as $p_{r|s}(r|s; \theta)$ or $p_{r|s}(r|s; \beta)$.
For example, a candidate model of $C$ can be decomposed as 
\begin{align*}
p_c(y, z, a; \beta) = p_x(y, z; \theta)p_{a|x}(a | y, z; \beta). 
\end{align*}

\subsection{Comparing the two estimators}

We have thus far obtained the two MLEs of $\theta$, namely $\hat\theta_y$ and $\hat\theta_b$, and their corresponding predictive distributions  $p_x(\hat{\theta}_y)$ and $p_x(\hat{\theta}_b)$, respectively.
We would like to determine which of the two predictive distributions approximates $q_x$ better than the other.
The approximation error of $p_x(\theta)$ is measured by the KL divergence from $q_x$ to $p_x(\theta)$ defined as
\begin{align*}
D_x(q_x; p_x(\theta))=-\int q_x(x) \log{p_x(x; \theta)}dx +\int q_x(x) \log{q_x(x)}dx. 
\end{align*}
Since the last term on the right hand side does not depend on $p_x(\theta)$, we ignore it for computing the loss function of $p_x(\theta)$ defined by
\begin{align*}
\mathcal{L}_x(\theta)=-\int q_x(x) \log{p_x(x; \theta)}dx. 
\end{align*}
Let $\hat{\theta}$ be an estimator of $\theta$. The risk (or expected loss) function of $p_x(\hat{\theta})$ is defined by
\begin{align}
\mathcal{R}_x(\hat{\theta})=E[\mathcal{L}_x(\hat{\theta})],
\label{risk}
\end{align}
where we take the expectation by considering $\hat{\theta}$ as a random variable.
Note that $\hat\theta$ in the notation of  $\mathcal{R}_x(\hat{\theta})$ indicates the procedure for computing $\hat\theta$ instead of a particular value of $\hat\theta$.
$\mathcal{R}_x(\hat{\theta})$ measures how well $p_x(\hat \theta)$ approximates $q_x$ on average in the long run.

For comparing the two MLEs, we define $\mathcal{R}_x(\hat{\theta}_y)$ and $\mathcal{R}_x(\hat{\theta}_b)$
by considering that $\hat{\theta}_y$ and $\hat{\theta}_b$ are functions of
independent random variables $Y_1,\ldots, Y_n$ and $B_1,\ldots, B_n$, respectively, where $B_i=(Y_i, A_i)$ has the same distribution as $B$ for all $i=1, \ldots, n$.
$\hat{\theta}_b$ is better than $\hat{\theta}_y$ when $\mathcal{R}_x(\hat{\theta}_b) < \mathcal{R}_x(\hat{\theta}_y)$, that is, the auxiliary variable $A$ helps the statistical inference on $q_x$.
On the other hand, $A$ is harmful when $\mathcal{R}_x(\hat{\theta}_b) > \mathcal{R}_x(\hat{\theta}_y)$.
Although we focus only on comparison between $Y$ and $B=(Y,A)$ in this paper, if there are more than two auxiliary variables (and their combinations) $A_1, A_2, \ldots$, then we may compare  $\mathcal{R}_x(\hat{\theta}_{(y,a_1)}), \mathcal{R}_x(\hat{\theta}_{(y,a_2)}), \ldots$, to determine good auxiliary variables.
Of course, the risk functions cannot be calculated in reality because they depend on the unknown true distribution. Thus, we derive a new information criterion as an estimator of the risk function in our setting. Since an asymptotically unbiased estimator of $\mathcal{R}_x(\hat{\theta}_y)$ has been already derived in \citet{SM18}, we will only derive an asymptotically unbiased estimator of $\mathcal{R}_x(\hat{\theta}_b)$. 

\section{An illustrative example with auxiliary variables}\label{ie}
\subsection{Model setting}

In this section, we demonstrate parameter estimation by using auxiliary variables in Gaussian mixture model (GMM), which can be formulated in incomplete data analysis. Let us consider two-component GMM; observed values are generated from one of two Gaussian distributions, where the assigned labels are missing. The observed data and missing labels are realizations of $Y$ and $Z$, respectively.
We estimate a predictive distribution of $X=(Y, Z)$ from the observation of $Y$, and we attempt improving it by utilizing $A$ in addition to $Y$. 
The true density function of primary variables $X=(Y, Z) \in \mathbb{R} \times \{0,1\}$ is given as 
\begin{align*}
q_{y|z}(y|z)&=zN(y; -1.2, 0.7) + (1-z)N(y; 1.2, 0.7), \\
q_{z}(z)&=0.6z + 0.4(1-z), 
\end{align*}
where $N(\cdot; \mu, \sigma^2)$ denotes the density function of $N(\mu, \sigma^2)$, i.e., the normal distribution with mean $\mu$ and variance $\sigma^2$.
We consider the following two cases for the true conditional distribution of auxiliary variable $A$ given $X=x$:
\begin{itemize}
\item[]
\begin{itemize}
\item[Case 1:] $q_{a|x}(a|y, z)=q_{a|z}(a|z)=zN(a; 1.8, 0.49)+(1-z)N(a; -1.8, 0.49)$. 
\item[Case 2:] $q_{a|x}(a|y, z)=q_{a}(a)=0.6N(a; 1.8, 0.49) + 0.4N(a; -1.8, 0.49)$. 
\end{itemize}
\end{itemize}
The random variables $X$ and $A$ are not independent in Case 1 whereas they are independent in Case 2. Hence, $A$ will contribute to estimating $\theta$ in Case 1. On the other hand, in Case 2, $A$ must not be useful, and $A$ becomes just noise if we estimate $\theta$ from $Y$ and $A$. 

In both cases, we use the following two-component GMM as a candidate model of $q_c$:
\begin{align}
\begin{split}
p_{b|z}(y, a | z; \beta)&=z N_2((y, a)^\top; \mu_1, \Sigma) + (1-z) N_2((y, a)^\top; \mu_2, \Sigma), \\
p_{z}(z; \theta)&=\pi_1z + (1-\pi_1)(1-z), 
\end{split}
\label{simu.1.pz}
\end{align}
where $N_2(\cdot ; \mu_i, \Sigma)$ denotes the density function of bivariate normal distribution $N_2(\mu_i, \Sigma)$, $i=1, 2$, and the parameters are
\begin{align*}
\mu_1=\begin{pmatrix}
\mu_{1y}\\
\mu_{1a}
\end{pmatrix}, ~~
\mu_2=\begin{pmatrix}
\mu_{2y}\\
\mu_{2a}
\end{pmatrix}, ~~ 
\Sigma=\begin{pmatrix}
\sigma_y^2 & \sigma_{ya}\\
\sigma_{ya} & \sigma_a^2
\end{pmatrix}. 
\end{align*} 
Therefore, $\beta=(\theta^\top, \varphi^\top)^\top$, $\theta=(\pi_1, \mu_{1y}, \mu_{2y}, \sigma_y^2)^\top$ and $\varphi=(\mu_{1a}, \mu_{2a}, \sigma_a^2, \sigma_{ya})^\top$.
The true parameters of $\theta$ and $\varphi$ for Case~1 are given by $\theta_0=(0.6, -1.2, 1.2, 0.7)^\top$ and $\varphi_0=(1.8, -1.8, 0.49, 0)^\top$, respectively.
By considering the joint density function $p_c(y,z,a; \beta) = p_{b|z}(y, a | z; \beta) p_{z}(z; \theta)$,
this candidate model correctly specifies the true density function $q_c(y,z,a)=q_{a|x}(a|y, z) q_{y|z}(y|z) q_{z}(z)$ in Case~1.
On the other hand, the model is misspecified for Case~2, and we cannot think of the true parameters.

\subsection{Estimation results} \label{sec:estimation_results}

For illustrating the impact of auxiliary variables on parameter estimation in each case, we generated a typical dataset $c_1,\ldots, c_n$ with sample size $n=100$ from $q_c$, which is actually picked from 10,000 datasets generated in the simulation study of Section \ref{simu}, and details of how to select the typical dataset are also shown there. For each case, we computed the three MLEs $\hat\theta_y$, $\hat\theta_b$ and $\hat\theta_x$, where $\hat\theta_x$ is the MLE of $\theta$ calculated by using complete data $x_1, \ldots, x_n$ as if labels $z_1, \ldots, z_n$ were available.

\begin{figure}[htbp]
\begin{center}
\begin{tabular}{c}
\begin{minipage}{0.49\hsize}
  \includegraphics[width=1\textwidth]{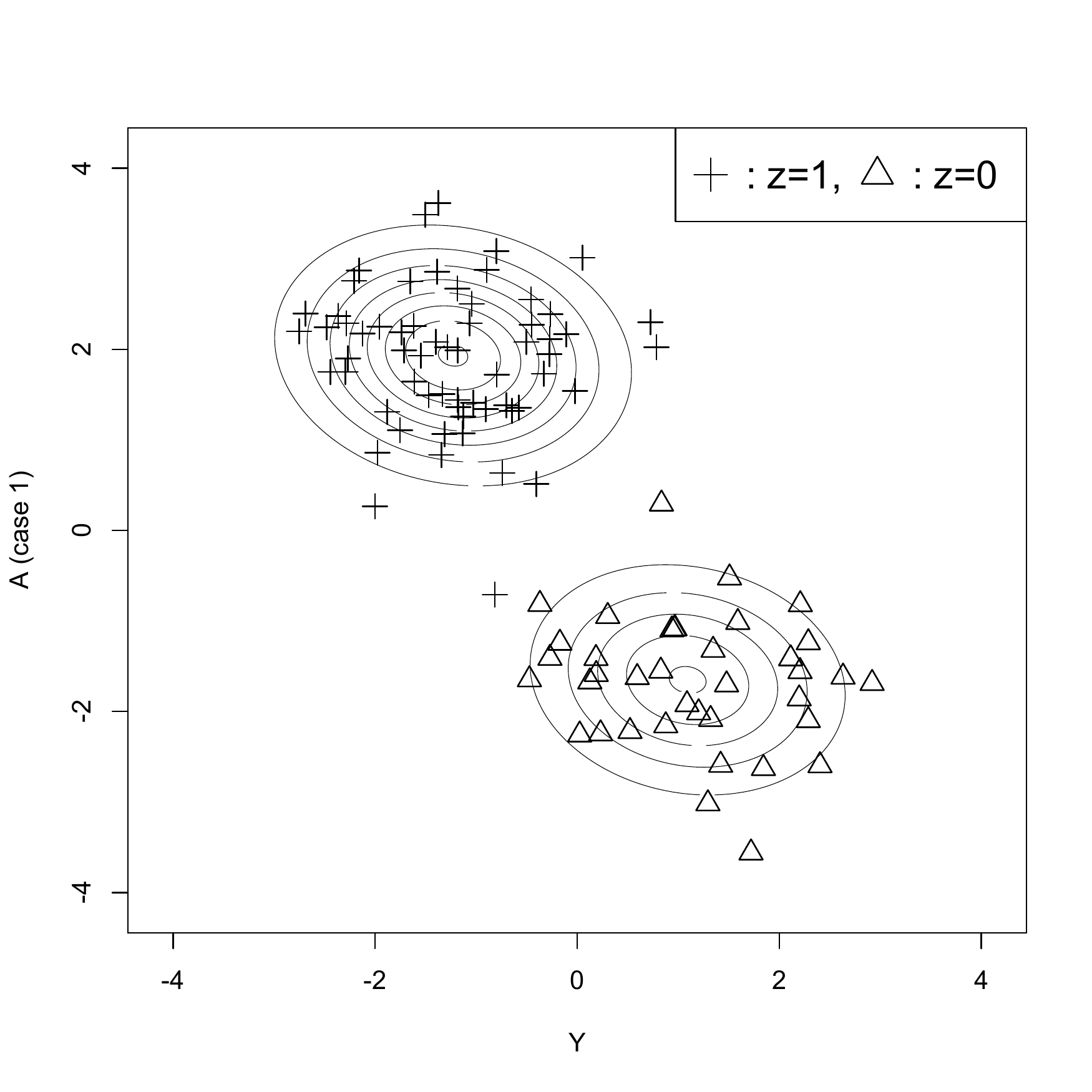}
\end{minipage}
\begin{minipage}{0.49\hsize}
  \includegraphics[width=1\textwidth]{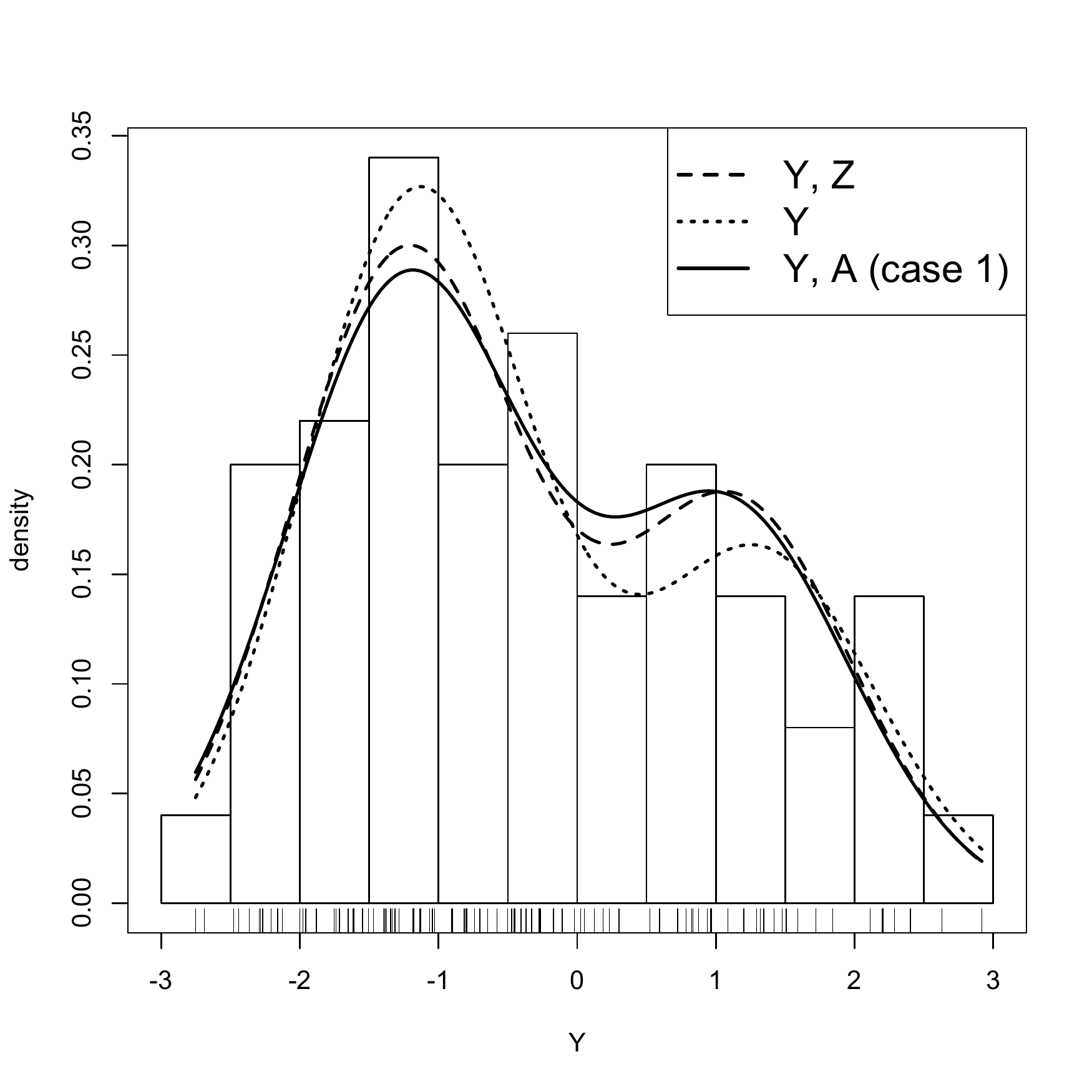}
\end{minipage}
\end{tabular}
\caption{{\it Useful} auxiliary variable (Case 1). 
The left panel plots $\{(y_i, a_i)\}_{i=1}^{100}$ with labels indicating $z_i$.
The estimated $p_b(\hat{\beta}_b)$ is shown by the contour lines.
The right panel shows the histogram of $\{y_i\}_{i=1}^{100}$, and three density functions $p_y(\hat\theta_x)$ (broken line), $p_y(\hat{\theta}_y)$ (dotted line) and $p_y(\hat{\theta}_b)$ (solid line).
In Section~\ref{sec:aic_comp}, this useful auxiliary variable is selected by our method (Case 1 in Table~\ref{tab:aic_comp}). 
}
\label{plot1}
\end{center}
\end{figure}

The result of Case~1 is shown in Figure~\ref{plot1}, where $A$ is beneficial for estimating $\theta$.
In the left panel, the two clusters are well separated, which makes parameter estimation stable.
The estimated $p_b(\hat\beta_b)$ captures the structure of the two clusters corresponding to the label $z_i=0$ and $z_i=1$, showing that  $p_c(\hat\beta_b)$ is estimated reasonably well, and thus $p_x(\hat\theta_b)$ is a good approximation of $q_x$.
Looking at the right panel, we also observe that
$p_y(\hat{\theta}_b)$ is better than $p_y(\hat{\theta}_y)$ for approximating $p_y(\hat\theta_x)$,
suggesting that the auxiliary variable is useful for recovering the lost information of missing data.
In fact, the three MLEs are calculated as follows: $\hat{\theta}_y=(0.671, -1.143, 1.324,  0.678)^\top$, $\hat{\theta}_b=(0.613, -1.228, 1.093, 0.744)^\top$ and $\hat{\theta}_x=(0.620, -1.233, 1.141, 0.695)^\top$. By comparing $\|\hat{\theta}_b- \hat{\theta}_x\|=0.069$ with $\|\hat{\theta}_y- \hat{\theta}_x\|=0.212$, we can see that $\hat{\theta}_b$ is better than $\hat{\theta}_y$ for predicting $\hat{\theta}_x$ without looking at the latent variable. 
All these observations indicate that the parameter estimation of $\theta$ is improved by using $A$ in Case~1.

\begin{figure}[htbp]
\begin{center}
\begin{tabular}{c}
\begin{minipage}{0.49\hsize}
  \includegraphics[width=1\textwidth]{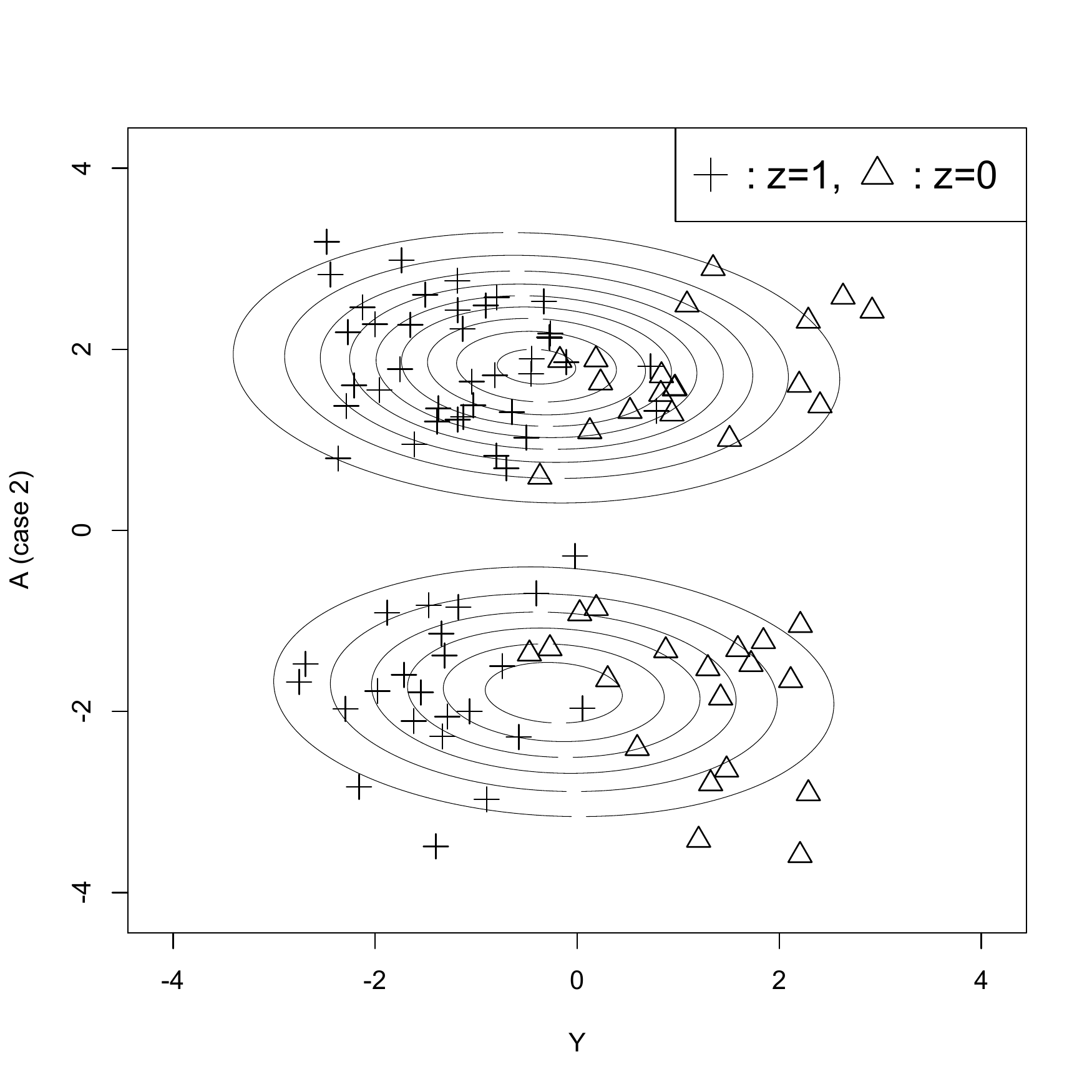}
\end{minipage}
\begin{minipage}{0.49\hsize}
  \includegraphics[width=1\textwidth]{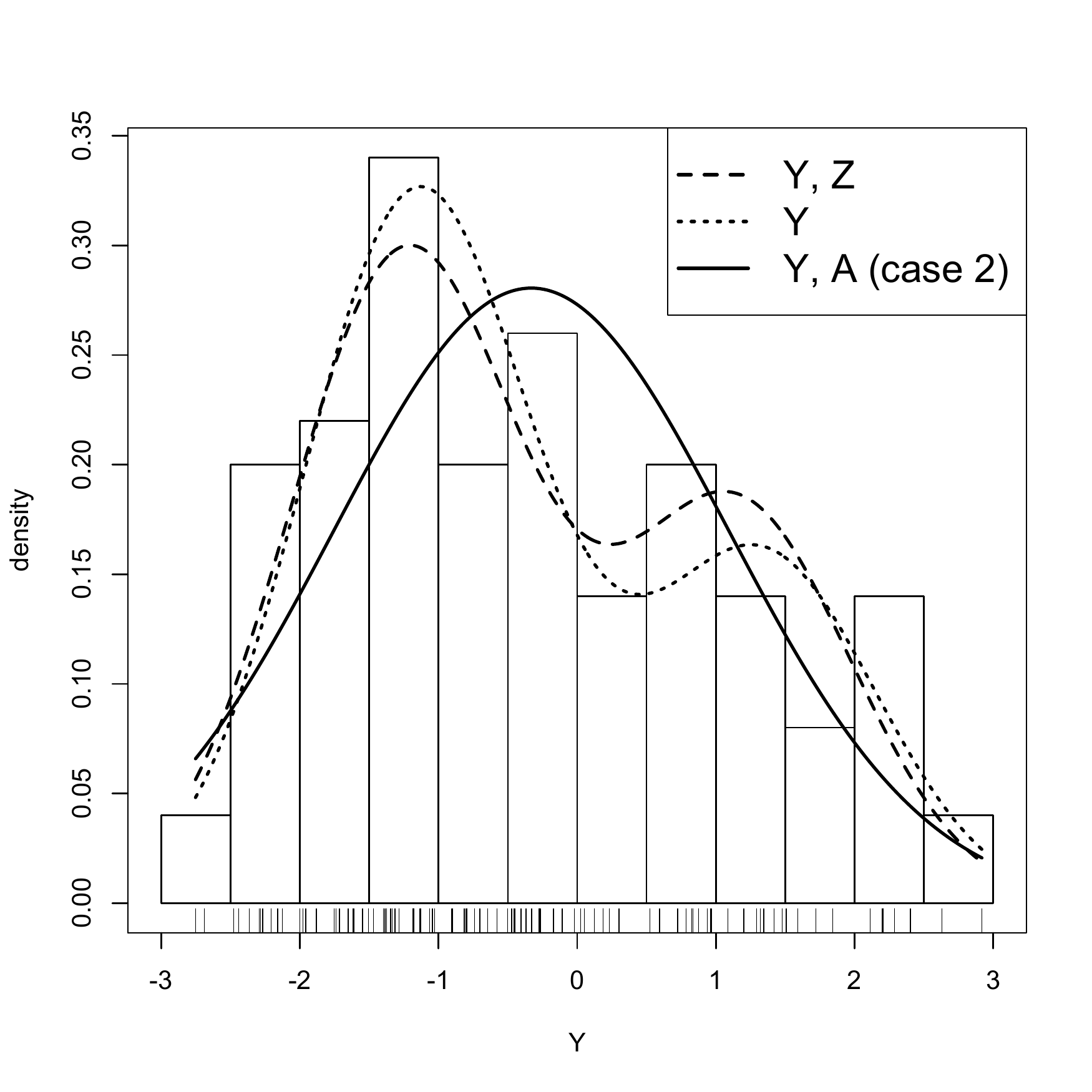}
\end{minipage}
\end{tabular}
\caption{{\it Useless} auxiliary variable (Case 2). 
The symbols are the same as Figure~\ref{plot1}.
In Section~\ref{sec:aic_comp}, this useless auxiliary variable is \emph{NOT} selected by our method (Case 2 in Table~\ref{tab:aic_comp}).
}
\label{plot2}
\end{center}
\end{figure}

The result of Case~2 is shown in Figure~\ref{plot2}, where $A$ is harmful for estimating $\theta$.
For fair comparison, exactly the same values of $\{(y_i, z_i)\}_{i=1}^{100}$ are used in both cases.
Thus, $\hat{\theta}_y$ and $\hat{\theta}_x$ have the same values as in Case 1 whereas $\hat{\theta}_b$ has a different value as $\hat{\theta}_b=(0.581, -0.403, -0.232, 2.015)^\top$.
By comparing $\|\hat{\theta}_b- \hat{\theta}_x\|=2.078$ with $\|\hat{\theta}_y- \hat{\theta}_x\|=0.212$, we can see that $\hat{\theta}_b$ is worse than $\hat{\theta}_y$ for predicting $\hat{\theta}_x$.
This is also seen in Figure~\ref{plot2}.
In the left panel, the estimated $p_b(\hat\beta_b)$ captures some structure of the two clusters, but they do not correspond to the label $z_i=0$ and $z_i=1$.
As a result, $p_y(\hat{\theta}_b)$ becomes a very poor approximation of $p_y(\hat\theta_x)$ in the right panel, indicating that the parameter estimation of $\theta$ is actually hindered by using $A$ in Case 2.

These examples suggest that usefulness of auxiliary variables depends strongly on the true distribution and a candidate model. Hence, it is important to select useful auxiliary variables from observed data.

\section{Information criterion}\label{ic}

\subsection{Asymptotic expansion of the risk function}

In this section, we derive a new information criterion as an asymptotically unbiased estimator of the risk function $\mathcal{R}_x(\hat{\theta}_b)$ defined in (\ref{risk}). 
We start from a general framework of misspecification, i.e., without assuming that candidate models are correctly specified, and later we give specific assumptions.
Let $\bar{\beta}$ be the optimal parameter value with respect to the KL divergence from $q_b$ to $p_b(\beta)$, that is, 
\begin{align*}
\bar{\beta}=\begin{pmatrix}
\bar{\theta}\\ 
\bar{\varphi}
\end{pmatrix}={\rm arg}\max_{\beta \in \mathcal{B}} \int q_b(b) \log p_b(b; \beta)d b. 
\end{align*}
If the candidate model is correctly specified, i.e., there exists $\beta_0=(\theta_0^\top, \varphi_0^\top)^\top$ such that $q_b=p_b(\beta_0)$, then $\bar{\beta}=\beta_0$ as well as $\bar{\theta}=\theta_0$.

In this paper, we assume the regularity conditions A1 to A6 of \citet{W82} for $q_b$ and $p_b(\beta)$ so that the MLE $\hat{\beta}_b$ has consistency and asymptotic normality. In particular, $\bar\beta$ is determined uniquely (i.e., identifiable) and is interior to $\mathcal{B}$.
We assume that $I_b$ and $J_b$ defined below are nonsingular in the neighbourhood of $\bar\beta$. Then \citet{W82} showed the asymptotic normality
as $n\to\infty$,
\begin{align} \label{eq:asymptotic-normality-beta-b}
\sqrt{n} (\hat\beta_b - \bar\beta)  \stackrel{d}{\to}   N_{d+f}(0, I_b^{-1} J_b I_b^{-1}  ),
\end{align}
where $I_b$ and $J_b$ are $(d+f)\times(d+f)$ matrices defined by using
$\nabla=\partial/\partial \beta$, $\nabla^\top=\partial/\partial \beta^\top$ and $\nabla^2=\partial^2/\partial \beta\partial \beta^\top$ as
\begin{align*}
I_b=-E[\nabla^2 \log{p_b}(b ; \bar{\beta})],\quad J_b=E[\nabla \log{p_b}(b; \bar{\beta})\nabla^\top \log{p_b}(b; \bar{\beta})].
\end{align*}
Note that we write derivatives by abbreviated forms, e.g., $\nabla^2 \log{p_b}(b ; \bar{\beta})$ means $\nabla^2 \log{p_b}(b ; \beta)|_{\beta=\bar{\beta}}$ and so on. 
In addition, we allow interchange of integrals and derivatives rather formally when working with models, although we actually need conditions for the models such as \citet{W82}. Moreover, the condition A7 of \citet{W82} is assumed in order to establish $I_b=J_b$ when considering a situation that the candidate model is correctly specified. 
We assume the above conditions throughout the paper without explicitly stated. 

Let us define three $(d+f)\times(d+f)$ matrices as
\begin{align*}
I_x=-E[\nabla^2 \log{p_x}(x; \bar{\theta})], \quad I_y=-E[\nabla^2 \log{p_y}(y; \bar{\theta})],\quad 
I_{z|y}=-E[\nabla^2 \log p_{z|y}(z|y; \bar{\theta})] = I_x - I_y,
\end{align*}
which will be used in the lemmas below.
Since the derivatives of $\log{p_x}(x; \theta)$ and $\log{p_y}(y; \theta)$ with respect to $\varphi$ is zero, the matrices become singular when $f>0$, but this is not a problem in our calculation. 
The following lemma shows that the dominant term of  $\mathcal{R}_x(\hat{\theta}_b)$ is $\mathcal{L}_x(\bar{\theta})$ and the remainder terms are of order $O(n^{-1})$, by noting that $\nabla ^\top \mathcal{L}_x(\bar{\theta}) = O(1)$ and $E[\hat{\beta}_b-\bar{\beta}] = O(n^{-1})$ in general.
The proof is given in Appendix \ref{pf_lem1}. 
\begin{lemma}\label{lem1}
The risk function $\mathcal{R}_x(\hat{\theta}_b)$ is expanded asymptotically as  
\begin{align*}
\mathcal{R}_x(\hat{\theta}_b)&=\mathcal{L}_x(\bar{\theta}) +\nabla ^\top \mathcal{L}_x(\bar{\theta})E[\hat{\beta}_b-\bar{\beta}] +\frac{1}{2n}{\rm tr}(I_x I_b^{-1}J_bI_b^{-1}) +o(n^{-1}).
\end{align*}
\end{lemma}
Just as a remark, the term $\nabla ^\top \mathcal{L}_x(\bar{\theta})E[\hat{\beta}_b-\bar{\beta}]  = O(n^{-1})$ above does not appear in the derivation of AIC or TIC, where $B=X$ and thus $\nabla ^\top \mathcal{L}_x(\bar{\theta}) = 0$. This term appears when the loss function for evaluation and that for estimation differ, for example, in the derivation of the information criterion under covariate shift; see $K_w^{[1] \top} b_w$ in eq.~(4.1) of \citet{S00}.

\subsection{Estimating the risk function}

For deriving an estimator of $\mathcal{R}_x(\hat{\theta}_b)$, we introduce an additional condition.
Let us assume that the candidate model is correctly specified for the latent part as
\begin{align}
q_{z|y}(z|y)=p_{z|y}(z|y; \bar{\theta}). 
\label{assumption}
\end{align}
This is the same condition as eq.~(14) of \citet{SM18} except that $\bar\theta$ is replaced by
\begin{align*}
\bar\theta_y = {\rm arg}\max_{\theta \in \Theta} \int q_y(y) \log p_y(y; \theta)dy.
\end{align*}
Since $Z$ is missing completely in our setting, we need such a condition to proceed further.
Although any method cannot detect misspecification of $p_{z|y}$ if $p_b$ is correctly specified, it is often the case that misspecification of $p_{z|y}$ leads to that of $p_b$, and thus it is detected indirectly as in Case~2 of Section~\ref{ie}.

Note that the symbol of $\bar\theta$ in our notation should have been $\bar\theta_b$, although we used $\bar\theta$ for simplicity, and there is also $\bar\theta_x$ defined similarly from $p_x(x; \theta)$.  They all differ each other with differences of order $O(1)$ in general, but $\bar\theta = \bar\theta_y = \bar\theta_x = \theta_0$ when $p_c(\beta)$ is correctly specified as $q_c = p_c(\beta_0)$.

Now we give the asymptotic expansion of $E[\ell_y(\hat{\theta}_b)]$, which shows that $-\ell_y(\hat{\theta}_b)$ can be used as an estimator of
$\mathcal{L}_x(\bar{\theta})$ but the asymptotic bias is of order $O(n^{-1})$.
\begin{lemma}\label{lem2}
Assume the condition (\ref{assumption}). Then, the expectation of the estimated log-likelihood $\ell_y(\hat{\theta}_b)$ can be expanded as 
\begin{align*}
E[\ell_y(\hat{\theta}_b)]&=-\mathcal{L}_x(\bar{\theta}) -C(q_x) -\nabla ^\top \mathcal{L}_x(\bar{\theta}) E[\hat{\beta}_b-\bar{\beta}]
 +\frac{1}{n}{\rm tr}(I_b^{-1}K_{b,y})-\frac{1}{2n}{\rm tr}(I_y I_b^{-1}J_bI_b^{-1}) +o(n^{-1}), 
\end{align*}
where $K_{b,y}=E[\nabla \log{p_b}(\bar{\beta})\nabla^\top \log{p_y}(\bar{\theta})]$
and $C(q_x) = \int q_x(x) \log{q_{z|y}}(z|y)dx$.
\end{lemma}
The proof of Lemma \ref{lem2} is given in Appendix \ref{pf_lem2}. 
By eliminating $\mathcal{L}_x(\bar{\theta})$ from the two expressions in Lemma~\ref{lem1} and Lemma~\ref{lem2}, and rearranging the formula, 
we get the following lemma, which plays a central role in deriving our information criterion.
\begin{lemma}\label{thm1}
Assume the condition (\ref{assumption}). Then, an expansion of the risk function $\mathcal{R}_x(\hat{\theta}_b)$ is given by
\begin{align}
\mathcal{R}_x(\hat{\theta}_b)&=-E[\ell_y(\hat{\theta}_b)]-C(q_x)+\frac{1}{n}{\rm tr}(I_b^{-1}K_{b,y})
+\frac{1}{2n}{\rm tr}(I_{z|y} I_b^{-1}J_bI_b^{-1})+o(n^{-1}).
\label{risk_exp}
\end{align}
\end{lemma}
We can ignore $C(q_x)$ for model selection, because it is a constant term which does not depend on the candidate model.
Thus, finally, we define an information criterion from the right hand side of (\ref{risk_exp}).
The following theorem is an immediate consequence of Lemma~\ref{thm1}.
\begin{theorem}\label{thm2}
Assume the condition (\ref{assumption}).
Let us define an information criterion as
\begin{align}
\widehat{{\rm risk}}_{x; b}&=-2n\ell_y(\hat{\theta}_b)+2{\rm tr}(I_b^{-1}K_{b, y}) +{\rm tr}(I_{z|y} I_b^{-1}J_bI_b^{-1}). 
\label{risk_est}
\end{align}
Then this criterion is an asymptotically unbiased estimator of $2n\mathcal{R}_x(\hat{\theta}_b)$ by ignoring the constant term $C(q_x)$.
\begin{align*}
E[ \widehat{{\rm risk}}_{x; b}     ] = 2n\mathcal{R}_x(\hat{\theta}_b) + 2n C(q_x) + o(1).
\end{align*}
\end{theorem}
Note that the subscript of $\widehat{{\rm risk}}_{x; b}$, $x;b$ is defined in accordance with \citet{SM18}; thus the former $x$ and the latter $b$ mean random variables used in evaluation and estimation, respectively. This criterion is an extension of TIC because when $X=B=Y$, $\widehat{{\rm risk}}_{x; b}$ coincides with TIC of \citet{T76} defined as follows: 
\begin{align*}
    {\rm TIC} = -2n\ell_y(\hat{\theta}_y)+2{\rm tr}(I_y^{-1}J_y). 
\end{align*}

\subsection{Akaike information criteria for auxiliary variable selection}

In actual use, $\widehat{{\rm risk}}_{x; b}$ may have a too complicated form. Thus, we derive a simpler information criterion by assuming the correctness of the candidate model like as AIC. 
\begin{theorem}\label{thm3}
Suppose $p_c(\beta)$ is correctly specified so that $q_c = p_c(\beta_0)$ for some $\beta_0\in\mathcal{B}$.
Then, we have
\begin{align} \label{eq:IbJb-KbyIy}
    J_b = I_b,\quad K_{b,y} = I_y,
\end{align}
and thus $\widehat{{\rm risk}}_{x; b}$ is rewritten as
\begin{align}
{\rm AIC}_{x; b}&=-2n\ell_y(\hat{\theta}_b)+ {\rm tr}(I_x I_b^{-1}) +{\rm tr}(I_y I_b^{-1}). 
\label{AICxb}
\end{align}
This criterion is an asymptotically unbiased estimator of $2n\mathcal{R}_x(\hat{\theta}_b)$ by ignoring the constant term $C(q_x)$.
\begin{align*}
E[ {\rm AIC}_{x; b}   ] = 2n\mathcal{R}_x(\hat{\theta}_b) + 2n C(q_x) + o(1).
\end{align*}
\end{theorem}
The proof is given in Appendix \ref{pf_thm3}. $I_x$, $I_y$ and $I_b$ are replaced by their consistent estimators in practical situations. 

The newly obtained criterion ${\rm AIC}_{x; b}$ is a generalization of AIC and some of its variants.
If $\theta$ is estimated by $\hat{\theta}_y$ instead of $\hat{\theta}_b$, we simply let $B=Y$ in the expression of ${\rm AIC}_{x; b}$ so that
we get ${\rm AIC}_{x; y}$ proposed by \citet{SM18}:
\begin{align}
{\rm AIC}_{x; y}=-2n\ell_y(\hat{\theta}_y)+{\rm tr}(I_x I_y^{-1}) + d. 
\label{AICxy}
\end{align}
Note that if $B=Y$, $I_y$ is not singular because $\beta=\theta$.
On the other hand, if there is no latent part, we simply let $X=Y$ in the expression of ${\rm AIC}_{x; b}$ so that we get
\begin{align}
{\rm AIC}_{y; b}&=-2n\ell_y(\hat{\theta}_b) +2{\rm tr}(I_y I_b^{-1}). 
\label{AICyb}
\end{align}
This can be used to select useful auxiliary variables in complete data analysis. Moreover, if $X=Y=B$, ${\rm AIC}_{x; b}$ reduces to the original AIC proposed by \citet{A74}: \begin{align}
{\rm AIC}_{y; y}=-2n\ell_y(\hat{\theta}_y)+2d. 
\label{AICyy}
\end{align}
It is worth mentioning that ${\rm tr}(I_{z|y} I_b^{-1})$ is interpreted as the additional penalty for the latent part:
\begin{align*}
{\rm AIC}_{x; b} - {\rm AIC}_{y; b} = {\rm tr}(I_x I_b^{-1}) - {\rm tr}(I_y I_b^{-1}) = {\rm tr}(I_{z|y} I_b^{-1}) \geq 0,
\end{align*}
which is also mentioned in eq.~(1) of \citet{SM18} for the case of $B=Y$.

\subsection{The illustrative example (cont.)} \label{sec:aic_comp}

Let us return to the problem of determining whether to use the auxiliary variables or not, that is, comparison between $p_x(\hat\theta_b)$ and $p_x(\hat\theta_y)$. By comparing ${\rm AIC}_{x; b}$ with ${\rm AIC}_{x; y}$, we can determine whether the vector of auxiliary variables $A$ is useful or useless. Thus, only when ${\rm AIC}_{x; b} < {\rm AIC}_{x; y}$, we conclude that $A$ is useful in order to estimate $\theta$ for predicting $X$.

Let us apply this procedure to the illustrative example in Section \ref{ie}.
The generalized AICs are computed for the two cases of the typical dataset, and the results are shown in Table~\ref{tab:aic_comp}. 
Looking at the value of ${\rm AIC}_{x; b}-{\rm AIC}_{x; y}$, it is negative for Case~1, concluding that the auxiliary variable is useful, and it is positive for Case~2, concluding that the auxiliary variable is useless.
According to the AIC values, therefore, we use the auxiliary variable of Case 1, but do not use the auxiliary variable of Case 2.
This decision agrees with the observations of Figures~\ref{plot1} and \ref{plot2} in Section~\ref{sec:estimation_results}.
Actually, the decision is correct, because the value of $\mathcal{R}_x(\hat\theta_b) - \mathcal{R}_x(\hat\theta_y)$ is negative for Case~1 and positive for Case~2 as will be seen in the simulation study of Section~\ref{sec:simulation_selection}.

We can also argue the usefulness of the auxiliary variable for predicting $Y$ instead of $X$, that is, comparison between $p_y(\hat\theta_b)$ and $p_y(\hat\theta_y)$. By comparing ${\rm AIC}_{y; b}$ with ${\rm AIC}_{y; y}$, we can determine whether $A$ is useful or useless for predicting $Y$.
Looking at the value of ${\rm AIC}_{y; b}-{\rm AIC}_{y; y}$ in Table~\ref{tab:aic_comp}, we make the same decision as that for $X$.

\begin{table}[ht]
\centering
\caption{Comparisons between $\hat\theta_b$ and $\hat\theta_y$ for predicting $X$, and that for $Y$.}
\label{tab:aic_comp}
\begin{tabular}{ccc}
  \hline
 & $p_x(\hat\theta_b)$ vs. $p_x(\hat\theta_y)$ &  $p_y(\hat\theta_b)$ vs. $p_y(\hat\theta_y)$\\
 \cline{2-3}
 & ${\rm AIC}_{x; b}-{\rm AIC}_{x; y}$ & ${\rm AIC}_{y; b}-{\rm AIC}_{y; y}$ \\ 
  \hline
Case 1 & -2.67 & -0.96 \\ 
Case 2 & 9.86 & 10.37 \\ 
   \hline
\end{tabular}
\end{table}

\section{Leave-one-out cross validation}\label{cv}

Variable selection by cross-validatory (CV) choice \citep{S74} is often applied to real data analysis due to its simplicity, although its computational burden is larger than that of information criteria; see \citet{AC10} for a recent review of cross-validation methods.
As shown in \citet{S77}, leave-one-out cross validation (LOOCV) is asymptotically equivalent to TIC. 
Because LOOCV does not require calculation of the information matrices of TIC, LOOCV is easier to use than TIC.
There are also some literature for improving LOOCV such as \citet{YTM06}, which gives a modification of LOOCV to reduce its bias by considering  maximum weighted log-likelihood estimation.
However, we focus on the result of \citet{S77} and extend it to our setting. 

In incomplete data analysis, LOOCV cannot be directly used because the loss function with respect to the complete data includes latent variables. Thus, we transform the loss function as follows: 
\begin{align*}
\mathcal{L}_x(\theta)&= -\int q_y(y) g(y; \theta) dy, 
\end{align*}
where $g(y; \theta)=\log p_y(y; \theta)  + f(y; \theta)$ and 
\begin{align*}
f(y; \theta) = \int q_{z|y}(z|y) \log p_{z|y}(z|y; \theta) dz. 
\end{align*}
Note that $f(y; \theta) = 0$ when $X=Y$.
Using the function $g(y;\theta)$, we then obtain the following LOOCV estimator of  the risk function $\mathcal{R}_x(\hat{\theta}_b)$.
\begin{align*}
\mathcal{L}_x^{\rm cv}(\hat{\theta}_b)&=-\frac{1}{n}\sum_{i=1}^ng(y_i; \hat{\theta}_b^{(-i)}), 
\end{align*}
where $\hat{\theta}_b^{(-i)}$ is the leave-out-out estimate of $\theta$ defined as
\begin{align*}
\hat{\beta}_b^{(-i)}&=\begin{pmatrix}
\hat{\theta}_b^{(-i)}\\
\hat{\varphi}_b^{(-i)}
\end{pmatrix}={\rm arg}\max_{\beta \in \mathcal{B}} \frac{1}{n}\sum_{j \neq i}^n \log{p_b}(b_j; \beta)={\rm arg}\max_{\beta \in \mathcal{B}} \left\{ \ell_b(\beta) - \frac{1}{n}\log p_b(b_i; \beta) \right\}. 
\end{align*}
We will show below in this section that $\mathcal{L}_x^{\rm cv}(\hat{\theta}_b)$ is asymptotically equivalent to  $\widehat{{\rm risk}}_{x; b}$. 
For implementing the LOOCV procedure with latent variables, however, we have to estimate $q_{z|y}(z|y)$ by $p_{z|y}(z|y, \hat{\theta}_b)$ in $f(y;\theta)$.
This introduces a bias to $\mathcal{L}_x^{\rm cv}(\hat{\theta}_b)$, and hence, information criteria are preferable to the LOOCV in incomplete data analysis. 

Let us show the asymptotic equivalence of $\mathcal{L}_x^{\rm cv}(\hat{\theta}_b)$ and  $\widehat{{\rm risk}}_{x; b}$ by assuming that we know the functional form of $f(y; \theta)$.
Noting that $\hat{\beta}_b^{(-i)}$ is a critical point of $\ell_b(\beta) - \log p_b(b_i; \beta)/n$, we have 
\begin{align*}
\nabla \ell_b(\hat{\beta}_b^{(-i)}) = \frac{1}{n}\nabla \log p_b(b_i; \hat{\beta}_b^{(-i)}) = O_p(n^{-1}). 
\end{align*}
By applying Taylor expansion to $\nabla \ell_b(\beta)$ around   $\beta=\hat{\beta}_b$, it follows from $\nabla \ell_b(\hat{\beta}_b)=0$ that
\begin{align}
\nabla^2 \ell_b(\tilde{\beta}_b^i)(\hat{\beta}_b^{(-i)}-\hat{\beta}_b) = \frac{1}{n} \nabla \log p_b(b_i; \hat{\beta}_b^{(-i)}), \label{thm4_1}
\end{align}
where $\tilde{\beta}_b^i$ lies between $\hat{\beta}_b^{(-i)}$ and $\hat{\beta}_b$. We can see from (\ref{thm4_1}) that $\hat{\beta}_b^{(-i)}-\hat{\beta}_b=O_p(n^{-1})$. 
Next, we regard $g(y_i; \theta)$ as a function of $\beta$ and apply Taylor expansion to it  around $\beta=\hat{\beta}_b$. Therefore, $g(y_i; \hat{\theta}_b^{(-i)})$ can be expressed as follows: 
\begin{align}
g(y_i; \hat{\theta}_b^{(-i)})&=g(y_i; \hat{\theta}_b) +\nabla^\top g(y_i; \tilde{\theta}_b^i)(\hat{\beta}_b^{(-i)}-\hat{\beta}_b), \label{thm4_2}
\end{align}
where $\tilde{\theta}_b^i$ lies between $\hat{\theta}_b^{(-i)}$ and $\hat{\theta}_b$ ($\tilde{\theta}_b^i$ does not corresponding to $\tilde{\beta}_b^i$).
Then we assume that
\begin{align}
\frac{1}{n}\sum_{i=1}^n \nabla^2 \ell_b(\tilde{\beta}_b^i)^{-1} \nabla \log p_b(b_i; \hat{\beta}_b^{(-i)})\nabla^\top g(y_i; \tilde{\theta}_b^i) &\overset{p}{\rightarrow} -I_b^{-1} E[\nabla \log p_b(b; \bar{\beta})\nabla^\top g(y; \bar{\theta})]. 
\label{assumptioins_s77}
\end{align}
By noting $\hat{\beta}_b^{(-i)} = \hat{\beta}_b + O_p(n^{-1})$, we have $\tilde{\beta}_b^i = \bar{\beta}+O_p(n^{-1/2})$ and $\tilde{\theta}_b^i = \bar\theta+O_p(n^{-1/2})$,
and thus (\ref{assumptioins_s77}) holds at least formally.
With the above setup, we show the following theorem. The proof is given in Appendix \ref{pf_thm4}. 
\begin{theorem}\label{thm4}
Suppose the same assumptions of Theorem \ref{thm2} and (\ref{assumptioins_s77}),
we have 
\begin{align}
2n\mathcal{L}_x^{\rm cv}(\hat{\theta}_b)=\widehat{{\rm risk}}_{x; b}-2\sum_{i=1}^nf(y_i; \bar{\theta}) + o_p(1). 
\label{loocv_aicxb}
\end{align}
\end{theorem}
Because the second term on the right-hand side of (\ref{loocv_aicxb}) does not depend on candidate models under the condition (\ref{assumption}), this theorem implies that $\mathcal{L}_x^{\rm cv}(\hat{\theta}_b)$ is asymptotically equivalent to $\widehat{{\rm risk}}_{x; b}$ except for the scaling and the constant term.
However, someone may wonder why $f(y;\theta)$ is included in $g(y; \theta)$ for comparing models of $p(b; \beta)$.
By assuming that $p_{z|y}(\theta)$ is correctly specified for $q_{z|y}$, $f(y;\bar\theta) = \int q_{z|y}(z|y) \log q_{z|y}(z|y)dz$ does not depend on the model anymore, so we may simply exclude $f(y;\theta)$ from $g(y;\theta)$, leading to the loss $\mathcal{L}_y(\theta)$ instead.
The reason of including $f(y;\theta)$ in $g(y;\theta)$ is explained as follows.
$\mathcal{L}_x^{\rm cv}(\hat{\theta}_b)$ as well as $\widehat{{\rm risk}}_{x; b}$ (and ${\rm AIC}_{x; b}$) includes the additional penalty for estimating $\hat\theta_b$ in $f(y; \hat\theta_b)$, which depends on the candidate models even if  $p_{z|y}(\theta)$ is correctly specified.

\section{Experiments with simulated datasets}\label{simu}

This section shows the usefulness of auxiliary variables and the proposed information criteria via a simulation study. The models illustrated in Section \ref{ie} are used for confirming the asymptotic unbiasedness of the information criterion and the validity of auxiliary variable selection.

\subsection{Unbiasedness}\label{unbiasedness}
At first, we confirm the asymptotic unbiasedness of ${\rm AIC}_{x; b}$ for estimating $2n\mathcal{R}_x(\hat{\theta}_b)$ except for the constant term, $C(q_x)$. The simulation setting is the same as Case 1 in Section \ref{ie}, thus the data generating model is given by
\begin{align*}
q_{b|z}(y, a|z)&=zN_2((y, a)^\top; \mu_{10}, \Sigma_0)
 + (1-z)N_2((y, a)^\top; \mu_{20}, \Sigma_0), \\
q_{z}(z)&=0.6z + 0.4(1-z), 
\end{align*}
where $\mu_{10}=-\mu_{20}=(-1.2, 1.8)^\top$ and $\Sigma_0={\rm diag}(0.7, 0.49)$. We generated $T=10^4$ independent replicates of the dataset $\{(y_i, z_i, a_i)\}_{i=1}^n$ from this model; in fact, we used $\{(y_i, z_i, a_{i,1})\}_{i=1}^n$ generated in Section~\ref{sec:simulation_selection}.
The candidate model is given by (\ref{simu.1.pz}), which is correctly specified for the above data generating model.
Because ${\rm AIC}_{x; b}$ is derived by ignoring $C(q_x)$, we compare $E[{\rm AIC}_{x; b}-{\rm AIC}_{x; y}]$ with $2n\{\mathcal{R}_x(\hat{\theta}_b) - \mathcal{R}_x(\hat{\theta}_y)\}$.
The computation of the expectation is approximated by the simulation average as
\begin{align*}
E[{\rm AIC}_{x; b}-{\rm AIC}_{x; y}]&\approx \frac{1}{T}\sum_{t=1}^{T} \{{\rm AIC}_{x; b}^{(t)}- {\rm AIC}_{x; y}^{(t)}\} , \\
2n\{\mathcal{R}_x(\hat{\theta}_b) - \mathcal{R}_x(\hat{\theta}_y)\}&\approx \frac{2n}{T}\sum_{t=1}^{T}\{\mathcal{L}_x(\hat{\theta}_b^{(t)}) -\mathcal{L}_x(\hat{\theta}_y^{(t)})\}, 
\end{align*}
where ${\rm AIC}_{x; b}^{(t)}$, ${\rm AIC}_{x; y}^{(t)}$, $\hat{\theta}_b^{(t)}$ and $\hat{\theta}_y^{(t)}$ are those computed for the $t$-th dataset ($t=1, \ldots, T$).

Here, we remark about calculation of the loss function $\mathcal{L}_x(\hat{\theta})$ in two-component GMM. Let $\hat{\theta}=(\hat{\pi}_1, \hat{\mu}_1, \hat{\mu}_2, \hat{\sigma}^2)^\top$ be an estimator of $\theta$. We expect that the components of GMM corresponding to $Z=1$ and $Z=0$  consists of $(\hat{\pi}_1, \hat{\mu}_1, \hat{\sigma}^2)$ and $(1-\hat{\pi}_1, \hat{\mu}_2, \hat{\sigma}^2)$, respectively. However, we cannot determine the assignment of the estimated parameters in reality, i.e., $(\hat{\pi}_1, \hat{\mu}_1, \hat{\sigma}^2)$ and $(1-\hat{\pi}_1, \hat{\mu}_2, \hat{\sigma}^2)$ may correspond to $Z=0$ and $Z=1$, respectively, because the labels $z_1, \ldots, z_n$ are missing. The assignment is required to calculate $\mathcal{L}_x(\hat{\theta})$ whereas it is not used for $\mathcal{L}_y(\hat{\theta})$ and the proposed information criteria. Hence, in this paper, we define $\mathcal{L}_x(\hat{\theta})$ as the minimum value between $\mathcal{L}(\hat{\theta})$ and $\mathcal{L}(\hat{\theta}')$, where $\hat{\theta}'=(1-\hat{\pi}_1, \hat{\mu}_2, \hat{\mu}_1, \hat{\sigma}^2)^\top$. 

Table \ref{simu1.1} shows the result of the simulation for $n=100, 200, 500, 1000, 2000$ and $5000$. For all $n$,  we observe that $E[{\rm AIC}_{x; b}-{\rm AIC}_{x; y}]$ is very close to  $2n\{\mathcal{R}_x(\hat{\theta}_b) - \mathcal{R}_x(\hat{\theta}_y)\}$, indicating the unbiasedness of ${\rm AIC}_{x; b}$.

\begin{table}[htbp]
\centering
\caption{Expected AIC difference is compared with the risk difference. The values are computed from $T=10^4$ runs of simulation with their standard errors in parentheses. }
\begin{tabular}{ccccccc}
  \hline
$n$ & 100 & 200 & 500 & 1000 & 2000 & 5000 \\ 
  \hline
$E[{\rm AIC}_{x; b}-{\rm AIC}_{x; y}]$  & -3.559 & -3.263 & -3.221 & -3.197 & -3.195 & -3.180 \\[-2ex]
 & (0.074) & (0.021) & (0.015) & (0.013) & (0.013) & (0.012) \\ 
$2n\{\mathcal{R}_x(\hat{\theta}_b) - \mathcal{R}_x(\hat{\theta}_y)\}$  & -3.603 & -3.333 & -3.275 & -3.208 & -3.182 & -3.232 \\[-2ex]
 & (0.071) & (0.054) & (0.050) & (0.050) & (0.050) & (0.050) \\ 
   \hline
\end{tabular}
\label{simu1.1}
\end{table}

\subsection{Auxiliary variable selection} \label{sec:simulation_selection}
Next, we demonstrate that the proposed AIC selects a useful auxiliary variable (Case 1), while it does not select a useless auxiliary variable (Case 2).
In each case, we generated $T=10^4$ independent replicates of the dataset $\{(y_i, z_i, a_i)\}_{i=1}^n$ from the model.
Actually, the values of $\{(y_i, z_i)\}_{i=1}^n$ are shared in both cases, so we generated replicates of $\{(y_i, z_i, a_{i,1}, a_{i,2} )\}_{i=1}^n$, where $a_{i,1}$ and $a_{i,2}$ are auxiliary variables for Case 1 and Case 2, respectively.
In each case, we compute ${\rm AIC}_{x;b}$ and ${\rm AIC}_{x;y}$, then we select $\hat\theta_b$ (i.e., selecting the auxiliary variable $A$) if ${\rm AIC}_{x;b} < {\rm AIC}_{x;y}$ and select  $\hat\theta_y$ (i.e., not selecting the auxiliary variable $A$) otherwise. The selected estimator is denoted as $\hat{\theta}_{best}$.
This experiment was repeated for $T=10^4$ times.
Note that the typical dataset in Section~\ref{ie} was picked from the generated datasets so that it has around the median value in each of $\mathcal{L}_x(\hat\theta_b)-\mathcal{L}_x(\hat\theta_y)$, $\mathcal{L}_y(\hat\theta_b)-\mathcal{L}_y(\hat\theta_y)$, ${\rm AIC}_{x;b} - {\rm AIC}_{x;y}$ and ${\rm AIC}_{y;b} - {\rm AIC}_{y;y}$ in both cases.

The selection frequencies are shown in Tables \ref{simu1.2.1F} and \ref{simu1.2.2F}.
We observe that, as expected, the useful auxiliary variable tends to be selected in Case 1, while the useless auxiliary variable tends to be not selected in Case 2.

For verifying the usefulness of the auxiliary variable in both cases, we computed the risk value
$\mathcal{R}_x(\hat{\theta})$ for $\hat{\theta}=\hat{\theta}_y$, $\hat{\theta}_b$ and $\hat{\theta}_{best}$. They are approximated by the simulation average as
\begin{align*}
\mathcal{R}_x(\hat{\theta}) \approx \frac{1}{T}\sum_{t=1}^{T}\mathcal{L}_x(\hat{\theta}^{(t)}). 
\end{align*} 
The results are shown in Tables \ref{simu1.2.1L} and \ref{simu1.2.2L}.
For easier comparisons, the values are the differences from $\mathcal{L}_x(\theta_0)$ with the true value $\theta_0$.
For all $n$, we observe that, as expected, $\mathcal{R}_x(\hat{\theta}_b) < \mathcal{R}_x(\hat{\theta}_y)$ in Case 1, and $\mathcal{R}_x(\hat{\theta}_b) > \mathcal{R}_x(\hat{\theta}_y)$ in Case 2.
In both cases, $\mathcal{R}_x(\hat{\theta}_{best})$ is close to $\min\{\mathcal{R}_x(\hat{\theta}_b), \mathcal{R}_x(\hat{\theta}_y) \} $, indicating that the variable selection is working well.

\begin{table}[htbp]
\centering
\caption{{\it Useful} auxiliary variable (Case 1): selection frequencies of $\hat{\theta}_b$  and $\hat{\theta}_y$. }
\begin{tabular}{ccccccc}
  \hline
$n$ & 100 & 200 & 500 & 1000 & 2000 & 5000 \\ 
  \hline
  $\hat{\theta}_b$ & 9230 & 9475 & 9649 & 9687 & 9711 & 9727 \\ 
  $\hat{\theta}_y$ & 770 & 525 & 351 & 313 & 289 & 273 \\ 
   \hline
\end{tabular}
\label{simu1.2.1F}
\end{table}

\begin{table}[htbp]
\centering
\caption{{\it Useless} auxiliary variable (Case 2): selection frequencies of $\hat{\theta}_b$  and $\hat{\theta}_y$. }
\begin{tabular}{ccccccc}
  \hline
$n$ & 100 & 200 & 500 & 1000 & 2000 & 5000 \\ 
  \hline
 $\hat{\theta}_b$ & 1508 & 212 & 1 & 0 & 0 & 0 \\ 
 $\hat{\theta}_y$ & 8492 & 9788 & 9999 & 10000 & 10000 & 10000 \\ 
   \hline
\end{tabular}
\label{simu1.2.2F}
\end{table}

\begin{table}[htbp]
\centering
\caption{{\it Useful} auxiliary variable (Case 1): estimated risk functions of $\hat{\theta}_b$, $\hat{\theta}_y$ and $\hat{\theta}_{best}$, and their standard errors in parenthesis}
\begin{tabular}{ccccccc}
  \hline
 $n$ & 100 & 200 & 500 & 1000 & 2000 & 5000 \\ 
  \hline
  $2n\{\mathcal{R}_x(\hat{\theta}_b) - \mathcal{L}_x(\theta_0)\}$ & 4.229 & 4.079 & 4.051 & 4.039 & 4.029 & 4.033 \\[-2ex]
& (0.032) & (0.030) & (0.029) & (0.028) & (0.029) & (0.028) \\ 
  $2n\{\mathcal{R}_x(\hat{\theta}_y) - \mathcal{L}_x(\theta_0)\}$ & 7.831 & 7.412 & 7.326 & 7.247 & 7.211 & 7.266 \\[-2ex]
& (0.078) & (0.061) & (0.058) & (0.058) & (0.058) & (0.058) \\ 
  $2n\{\mathcal{R}_x(\hat{\theta}_{best}) - \mathcal{L}_x(\theta_0)\}$ & 5.109 & 4.741 & 4.501 & 4.491 & 4.479 & 4.454 \\[-2ex]
& (0.052) & (0.045) & (0.041) & (0.042) & (0.042) & (0.041) \\ 
   \hline
\end{tabular}
\label{simu1.2.1L}
\end{table}

\begin{table}[htbp]
\centering
\caption{{\it Useless} auxiliary variable (Case 2): estimated risk functions of $\hat{\theta}_b$, $\hat{\theta}_y$ and $\hat{\theta}_{best}$, and their standard errors in parenthesis}
\begin{tabular}{ccccccc}
  \hline
 $n$ & 100 & 200 & 500 & 1000 & 2000 & 5000 \\ 
  \hline
  $2n\{\mathcal{R}_x(\hat{\theta}_b) - \mathcal{L}_x(\theta_0)\}$ & 105.527 & 214.659 & 543.685 & 1091.105 & 2182.647 & 5452.623 \\[-2ex] 
& (0.111) & (0.167) & (0.301) & (0.474) & (0.723) & (1.151) \\ 
  $2n\{\mathcal{R}_x(\hat{\theta}_y) - \mathcal{L}_x(\theta_0)\}$ & 7.831 & 7.412 & 7.326 & 7.247 & 7.211 & 7.266 \\[-2ex] 
& (0.078) & (0.061) & (0.058) & (0.058) & (0.058) & (0.058) \\ 
  $2n\{\mathcal{R}_x(\hat{\theta}_{best}) - \mathcal{L}_x(\theta_0)\}$ & 22.064 & 11.555 & 7.375 & 7.247 & 7.211 & 7.266 \\[-2ex] 
& (0.358) & (0.304) & (0.079) & (0.058) & (0.058) & (0.058) \\ 
   \hline
\end{tabular}
\label{simu1.2.2L}
\end{table}

\section{Experiments with a real dataset}\label{sec:rd}

We show an example of auxiliary variable selection  using \emph{Wine Data Set} available at UCI Machine Learning Repository \citep{DT17}, which consists of 1 categorical variable (3 categories) and 13 continuous variables, denoted as $V_1, \ldots, V_{13}$.
For simplicity, we only use the first two categories and regard them as a latent variable $Z \in \{0, 1\}$; the experiment results were similar to the other combinations. The sample size is then $n=130$ and all variables except for $Z$ are standardized.
We set one of the 13 continuous variables as the observed primary variable $Y$, and set the rest of 12 variables as auxiliary variables $A_1, \ldots, A_{12}$.
For example, if $Y$ is $V_1$, then $A_1, \ldots, A_{12}$ are $V_2, \ldots, V_{13}$. 
The dataset is now $\{(y_i, z_i, a_{i,1}, \ldots, a_{i,12})\}_{i=1}^n$, which is randomly divided into the training set with sample size $n_{tr}=86$ ($z_i$ is not used) and the test set with sample size $n_{te}=44$ ($a_{i,1}, \ldots, a_{i,12}$ are not used).

In the experiment, we compute ${\rm AIC}_{x;b_\ell}$ for $B_\ell=(Y,A_\ell)$, $\ell=1,\ldots,12$, and ${\rm AIC}_{x;y}$ for $Y$ from the training dataset using the model (\ref{simu.1.pz}).
We select $\hat\theta_{best}$ from $\hat{\theta}_{b_1}, \ldots, \hat{\theta}_{b_{12}}$ and $\hat{\theta}_y$ by finding the minimum of the 13 AIC values. Thus we are selecting one of the auxiliary variables $A_1,\ldots,A_{12}$ or not selecting any of them.
It is possible to select a combination of the auxiliary variables, but we did not attempt such an experiment.
For measuring the generalization error, we compute $\mathcal{L}_x(\hat{\theta}_y) - \mathcal{L}_x(\hat{\theta}_{best})$ from the test set as
\begin{align*}
\mathcal{L}_x(\hat{\theta}_y) - \mathcal{L}_x(\hat{\theta}_{best}) \approx -\frac{1}{n_{te}}\sum_{i \in \mathcal{D}^{te}}
\{\log p_x(y_i, z_i; \hat{\theta}_y)
- \log p_x(y_i, z_i; \hat{\theta}_{best})\},
\end{align*}
where $\mathcal{D}^{te}\subset\{1, \ldots, n\}$ represents the test set.
The assignment problem of $\mathcal{L}_x(\cdot)$ mentioned in Section \ref{simu} is avoided by a similar manner. 

For each case of $Y=V_\ell$, $\ell=1, \ldots, 13$, the above experiment was repeated $100$ times, and the experiment average of the generalization error was computed.
The result is shown in Table~\ref{RD}.
A positive value indicates that $\hat\theta_{best}$ performed better than $\hat\theta_y$.
We observe that $\hat\theta_{best}$ is better than or almost the same as $\hat{\theta}_y$ for all cases $\ell=1, \ldots, 13$, suggesting that AIC works well to select a useful auxiliary variable.

\begin{table}[ht]
\centering
\caption{Experiment average of $n_{te}\{\mathcal{L}(\hat{\theta}_y) - \mathcal{L}_x(\hat{\theta}_{best})\}$ for each case of $Y=V_{\ell}$, $\ell=1, \ldots, 13$. Standard errors are in parenthesis.}
\begin{tabular}{cccccccc}
  \hline
$Y$ & $V_1$ & $V_2$ & $V_3$ & $V_4$ & $V_5$ & $V_6$ & $V_7$\\ 
  \hline
  $n_{te}\{\mathcal{L}_x(\hat{\theta}_y) - \mathcal{L}_x(\hat{\theta}_{best})\}$  & 0.13 & -0.14 & 89.71 & 46.24 & -1.76 & 3.34 & 76.54 \\[-2ex] 
   & (0.08) & (0.12) & (3.82) & (4.17) & (2.52) & (1.34) & (6.09) \\ 
  \hline
$Y$ & $V_8$ & $V_9$ & $V_{10}$ & $V_{11}$ & $V_{12}$ & $V_{13}$ \\ 
  \hline
  $n_{te}\{\mathcal{L}_x(\hat{\theta}_y) - \mathcal{L}_x(\hat{\theta}_{best})\}$ & 13.91 & 39.45 & 1.72 & 111.24 & 15.48 & 0.23 \\[-2ex] 
  & (2.21) & (3.12) & (0.29) & (8.46) & (2.11) & (0.09)\\
   \hline
\end{tabular}
\label{RD}
\end{table}

\section{Conclusion}\label{conc}

We often encounter a dataset composed of various variables. If only some of the variables are of interest, then the rest of the variables can be interpreted as auxiliary variables. Auxiliary variables may be able to improve estimation accuracy of unknown parameters but they could also be harmful.
Hence, it is important to select useful auxiliary variables. 

In this paper, we focused on exploiting auxiliary variables in incomplete data analysis. Usefulness of auxiliary variables is measured by a risk function based on the KL divergence for complete data. We derived an information criterion which is an asymptotically unbiased estimator of the risk function except for a constant term. Moreover, we extended a result of \citet{S77} to our setting and proved asymptotic equivalence between a variant of LOOCV and the proposed criteria. Since LOOCV requires an additional condition for its justification, the proposed criteria are preferable to LOOCV. 

This study assumes that variables are different between training set and test set.
There are other settings such as covariate shift \citep{S00} and transfer learning  \citep{PY10}, where distributions are different between training set and test set.
It will be possible to combine these settings to construct a generalized framework.
It is also possible to extend our study for taking account of a missing mechanism.
We will leave these extensions as future works.

\appendix\label{app}
\section{Proofs}
\subsection{Proof of Lemma \ref{lem1}}\label{pf_lem1}

\begin{proof}
Taylor expansion of $\mathcal{L}_x(\theta)$ around $\theta=\bar\theta$, by formally taking it as a function of $\beta$, gives
\begin{align*}
\mathcal{L}_x(\hat{\theta}_b)&=\mathcal{L}_x(\bar{\theta})+\nabla ^\top \mathcal{L}_x(\bar{\theta}) (\hat{\beta}_b-\bar{\beta})+\frac{1}{2}{\rm tr}\{I_x(\hat{\beta}_b-\bar{\beta})(\hat{\beta}_b-\bar{\beta})^\top\} + o_p(n^{-1}),
\end{align*}
where $\nabla^2 \mathcal{L}_x(\bar{\theta})=I_x$ is used above.
By taking the expectation of both sides, 
\begin{align*}
E[\mathcal{L}_x(\hat{\theta}_b)] &=\mathcal{L}_x(\bar{\theta})+\nabla ^\top \mathcal{L}_x(\bar{\theta})  E[\hat{\beta}_b-\bar{\beta}]+\frac{1}{2}{\rm tr}\{ I_{x} E[(\hat{\beta}_b-\bar{\beta})(\hat{\beta}_b-\bar{\beta})^\top]\} + o(n^{-1}) \\
&=\mathcal{L}_x(\bar{\theta}) +\nabla ^\top \mathcal{L}_x(\bar{\theta})  E[\hat{\beta}_b-\bar{\beta}]+\frac{1}{2n}{\rm tr}(I_x I_b^{-1}J_bI_b^{-1}) +o(n^{-1}),
\end{align*}
where the asymptotic variance of $\hat\beta_b$ in (\ref{eq:asymptotic-normality-beta-b}) is given as
\begin{align}
nE[(\hat{\beta}_b-\bar{\beta})(\hat{\beta}_b-\bar{\beta})^\top]=I_b^{-1}J_bI_b^{-1}+o(1). 
\label{theta_var}
\end{align}
\end{proof}

\subsection{Proof of Lemma \ref{lem2}}\label{pf_lem2}

\begin{proof}
Taylor expansion of $\ell_y(\theta)$ around $\theta=\bar\theta$, by formally taking it as a function of $\beta$, gives
\begin{align*}
\begin{split}
\ell_y(\hat{\theta}_b)=\ell_y(\bar{\theta}) +\nabla^\top \ell_y(\bar{\theta})(\hat{\beta}_b-\bar{\beta})- \frac{1}{2}{\rm tr}\{I_y (\hat{\beta}_b-\bar{\beta})(\hat{\beta}_b-\bar{\beta})^\top\} +o_p(n^{-1}),
\end{split}
\end{align*}
where $\nabla^2 \ell_y(\bar{\theta})=-I_y+o_p(1)$ is used above.
By taking the expectation of both sides,
\begin{align}
E[\ell_y(\hat{\theta}_b)]&=E[\ell_y(\bar{\theta})] +E[\nabla^\top \ell_y(\bar{\theta})(\hat{\beta}_b-\bar{\beta})]
-\frac{1}{2}E[{\rm tr}\{I_y (\hat{\beta}_b-\bar{\beta})(\hat{\beta}_b-\bar{\beta})^\top\}] +o(n^{-1})\nonumber \\
&=E[\ell_y(\bar{\theta})] +E[\nabla^\top \ell_y(\bar{\theta})(\hat{\beta}_b-\bar{\beta})] -\frac{1}{2n}{\rm tr}(I_yI_b^{-1}J_bI_b^{-1}) +o(n^{-1}).
\label{Lem2_expand1}
\end{align}
In the last expression, we used (\ref{theta_var}) for the asymptotic variance of $\hat\beta_b$.
For working on the second term in (\ref{Lem2_expand1}), we first derive an expression of $\hat{\beta}_b-\bar{\beta}$.
Taylor expansion of the score function $\nabla \ell_b(\beta)$ around $\beta=\bar\beta$ gives
\begin{align*}
\nabla \ell_b(\hat{\beta}_b) & = \nabla \ell_b(\bar\beta) + \nabla^2 \ell_b(\bar{\beta})(\hat{\beta}_b-\bar{\beta})+o_p(n^{-1/2})\\
& = \nabla \ell_b(\bar{\beta})-I_b(\hat{\beta}_b-\bar{\beta})+o_p(n^{-1/2}),
\end{align*}
where $\nabla^2\ell_b(\bar{\beta}) = - I_b+o_p(1)$ is used above.
By noticing $\nabla \ell_b(\hat{\beta}_b)=0$, we thus obtain
\begin{align} \label{eq:beta-by-score}
\hat{\beta}_b-\bar{\beta}=I_b^{-1}\nabla \ell_b(\bar{\beta})+o_p(n^{-1/2})=\frac{1}{n}\sum_{i=1}^nI_b^{-1}\nabla \log{p_b}(b_i; \bar{\beta})+o_p(n^{-1/2}),
\end{align}
where $E[\nabla \ell_b(\bar{\beta})]=0$ and each term in the summation has mean zero, because $E[\nabla \log{p_b}(b; \bar{\beta})]=\nabla E[\log{p_b}(b; \bar{\beta})]=0$.
Now we are back to the the second term in (\ref{Lem2_expand1}).
Using (\ref{eq:beta-by-score}), we have
\begin{align}
\nabla^\top \ell_y(\bar{\theta})(\hat{\beta}_b-\bar{\beta})&=E[\nabla^\top \ell_y(\bar{\theta})](\hat{\beta}_b-\bar{\beta}) +\{\nabla^\top \ell_y(\bar{\theta})-E[\nabla^\top \ell_y(\bar{\theta})]\}(\hat{\beta}_b-\bar{\beta}) \nonumber\\
&=E[\nabla ^\top \ell_y(\bar{\theta}) ](\hat{\beta}_b-\bar{\beta})+\{\nabla^\top \ell_y(\bar{\theta})-E[\nabla^\top \ell_y(\bar{\theta})]\}I_b^{-1}\nabla \ell_b(\bar{\beta})+o_p(n^{-1}). 
\label{Lem2_2nd_term1}
\end{align}
By noting $E[\nabla \ell_b(\bar{\beta})]=0$, the expectation of the second term in (\ref{Lem2_2nd_term1}) is
\begin{align}
E[\{\nabla^\top \ell_y(\bar{\theta})-E[\nabla^\top \ell_y(\bar{\theta})]\}I_b^{-1}\nabla \ell_b(\bar{\beta})]&=E[\nabla^\top \ell_y(\bar{\theta})I_b^{-1}\nabla \ell_b(\bar{\beta})]\nonumber\\
&=\frac{1}{n^2}\sum_{i=1}^n \sum_{j=1}^n E[\nabla^\top \log{p_y}(y_i; \bar{\theta})I_b^{-1}\nabla \log{p_b}(b_j; \bar{\beta})]\nonumber\\
&=\frac{1}{n}E[\nabla^\top \log{p_y}(y; \bar{\theta})I_b^{-1}\nabla \log{p_b}(b; \bar{\beta})] \nonumber \\
&=\frac{1}{n}{\rm tr}\{I_b^{-1}E[\nabla \log{p_b}(b; \bar{\beta})\nabla^\top \log{p_y}(y; \bar{\theta})]\} \nonumber \\
&=\frac{1}{n}{\rm tr}(I_b^{-1}K_{b,y}). 
\label{Lem2_2nd_term11}
\end{align}
Combining (\ref{Lem2_2nd_term1}) and (\ref{Lem2_2nd_term11}), we have
\begin{align}
E[\nabla^\top \ell_y(\bar{\theta})(\hat{\beta}_b-\bar{\beta})]=E[\nabla ^\top \ell_y(\bar{\theta}) ] E[\hat{\beta}_b-\bar{\beta}]+\frac{1}{n}{\rm tr}(I_b^{-1}K_{b,y})+o(n^{-1}). 
\label{Lem2_2nd_term2}
\end{align}
We next show that $E[\nabla ^ \top \ell_y(\bar{\theta}) ] =  -\nabla ^\top \mathcal{L}_x(\bar{\theta})$.
Let us recall that we have assumed $q_{z|y}(z|y)=p_{z|y}(z|y; \bar{\theta})$ in (\ref{assumption}), which leads to
\begin{align*}
 E[ \nabla \log p_{z|y}(z|y; \bar\theta)] &= \int q_y(y) \int p_{z|y}(z|y ; \bar\theta) \nabla \log p_{z|y}(z|y; \bar\theta)\,dz dy\\
 &= \int q_y(y) \int \nabla  p_{z|y}(z|y; \bar\theta)\,dz dy\\
 &= \int q_y(y) \nabla \!\int p_{z|y}(z|y; \bar\theta)\,dz dy = 0.
 \end{align*}
Therefore,
\begin{align*}
-\nabla \mathcal{L}_x(\bar{\theta}) &= \nabla E[\log{p_x(x; \bar{\theta})}]\\
&=E[\nabla \log{p_x(x; \bar{\theta})}]\\
&=E[\nabla \log{p_y(y; \bar{\theta})}] + E[\nabla \log{p_{z|y}(z|y; \bar{\theta})}]\\
&=E[\nabla \ell_y(\bar{\theta})]. 
\end{align*}
Substituting this and (\ref{Lem2_2nd_term2}) into the second term in (\ref{Lem2_expand1}), we have
\begin{align}
E[\ell_y(\hat{\theta}_b)]&=E[\ell_y(\bar{\theta})] -\nabla ^\top \mathcal{L}_x(\bar{\theta}) E[\hat{\beta}_b-\bar{\beta}] \nonumber \\
&~~~~ +\frac{1}{n}{\rm tr}(I_b^{-1}K_{b,y})-\frac{1}{2n}{\rm tr}(I_yI_b^{-1}J_bI_b^{-1}) +o(n^{-1}). \label{Lem2_afterinsert}
\end{align}
The first term on the right hand side in (\ref{Lem2_afterinsert}) is
\begin{align*}
E[\ell_y(\bar{\theta})] &= E[\log{p_y}(y; \bar{\theta}) ]\\
&=E[\log{p_x}(x; \bar{\theta}) ] - E[\log p_{z|y}(z|y; \bar{\theta})]\\
&=-\mathcal{L}_x(\bar{\theta}) - C(q_x),
\end{align*}
where  (\ref{assumption}) is used again in the last term.
Finally (\ref{Lem2_afterinsert}) is rewritten as
\begin{align*}
E[\ell_y(\hat{\theta}_b)]&=-\mathcal{L}_x(\bar{\theta}) -C(q_x)-\nabla ^\top \mathcal{L}_x(\bar{\theta}) E[\hat{\beta}_b-\bar{\beta}] \\
&~~~~ +\frac{1}{n}{\rm tr}(I_b^{-1}K_{b,y})-\frac{1}{2n}{\rm tr}(I_yI_b^{-1}J_bI_b^{-1}) +o(n^{-1}). 
\end{align*}
\end{proof}

\subsection{Proof of Theorem \ref{thm3}}\label{pf_thm3}

\begin{proof}
First recall that we have assumed that $q_c(c) = p_c(c; \beta_0)$, which also implies the condition (\ref{assumption})
as $q_{z|y}(z|y) = p_{z|y}(z|y ; \theta_0)$ with $\bar\beta = \beta_0$. Thus Theorem~\ref{thm2} holds.
Substituting $J_b = I_b$ and $K_{b,y}=I_y$ in the penalty term of (\ref{risk_est}), we have
\begin{align*}
2{\rm tr}(I_b^{-1}K_{b, y}) +{\rm tr}(I_{z|y} I_b^{-1}J_bI_b^{-1})
= 2{\rm tr}(I_b^{-1}I_y) +{\rm tr}((I_x - I_y) I_b^{-1}) 
=  {\rm tr}(I_b^{-1}I_y) + {\rm tr}(I_x I_b^{-1}),
\end{align*}
giving the penalty term of (\ref{AICxb}). Therefore, we only have to show (\ref{eq:IbJb-KbyIy}). 
Noting the identity
\begin{align*}
   \nabla^2 \log p_b(b; \beta) = \frac{1}{p_b(b; \beta)} \nabla^2 p_b(b; \beta)  - \nabla \log p_b(b; \beta)  \nabla^\top \log p_b(b; \beta),
\end{align*}
it follows from $q_b(b) = p_b(b ; \beta_0)$ that
\begin{align*}
    I_b  = -E[\nabla^2 \log p_b(b; \beta_0)] &= -\int \nabla^2 p_b(b;\beta_0) db  + E[ \nabla \log p_b(b; \beta_0)  \nabla^\top \log p_b(b; \beta_0) ]\\
     &=- \nabla^2 \int  p_b(b;\beta_0) db + J_b = J_b.
\end{align*}
Note that the same result can be obtained from Theorem 3.3 in \citet{W82}.
Next we show $K_{b,y}=I_y$.
Since $q_{a|y}(a|y) = p_{a|y}(a|y; \beta_0)$, 
\begin{align*}
\int q_{a|y}(a|y) \nabla \log p_{a|y}(a|y ; \beta_0) da = \int \nabla  p_{a|y}(a|y ; \beta_0) da = \nabla \int   p_{a|y}(a|y ; \beta_0) da =0.
\end{align*}
Therefore, we have
\begin{align*}
    K_{b,y} &= E[ \nabla \log p_b(b; \beta_0) \nabla^\top \log p_y(y; \theta_0)] \\
      & = E[ \nabla \log p_y(y; \theta_0) \nabla^\top \log p_y(y; \theta_0)] + E[ \nabla \log p_{a|y}(a|y; \beta_0) \nabla^\top \log p_y(y; \theta_0)]\\
      & = I_y + \int q_y(y)  \left( \int q_{a|y}(a|y) \nabla \log p_{a|y}(a|y ; \beta_0) da \right) \nabla^\top \log p_y(y; \theta_0) dy\\
      &=I_y.
\end{align*}
\end{proof}

\subsection{Proof of Theorem \ref{thm4}}\label{pf_thm4}
\begin{proof}
It follows from (\ref{thm4_1}) and (\ref{thm4_2}) that 
\begin{align*}
g(y_i; \hat{\theta}_b^{(-i)})&=g(y_i; \hat{\theta}_b) +\frac{1}{n}\nabla^\top g(y_i; \tilde{\theta}_b^i)\nabla^2 \ell_b(\tilde{\beta}_b^i)^{-1}\nabla \log p_b(b_i; \hat{\beta}_b^{(-i)})\\
&=g(y_i; \hat{\theta}_b) +\frac{1}{n}{\rm tr}\{\nabla^2 \ell_b(\tilde{\beta}_b^i)^{-1}\nabla \log p_b(b_i; \hat{\beta}_b^{(-i)})\nabla^\top g(y_i; \tilde{\theta}_b^i)\}. 
\end{align*}
This and the assumption (\ref{assumptioins_s77}) imply that
\begin{align*}
\mathcal{L}_x^{\rm cv}(\hat{\theta}_b)&=-\frac{1}{n}\sum_{i=1}^ng(y_i; \hat{\theta}_b) -\frac{1}{n^2}\sum_{i=1}^n{\rm tr}\{\nabla^2 \ell_b(\tilde{\beta}_b^i)^{-1}\nabla \log p_b(b_i; \hat{\beta}_b^{(-i)})\nabla^\top g(y_i; \tilde{\theta}_b^i)\}\\
&=-\frac{1}{n}\sum_{i=1}^ng(y_i; \hat{\theta}_b)+\frac{1}{n}{\rm tr}\{I_b^{-1} E[\nabla \log p_b(\bar{\beta})\nabla^\top g(y; \bar{\theta})]\} + o_p(n^{-1}). 
\end{align*}
Under the assumption $q_{z|y}(z|y)=p_{z|y}(z|y; \bar{\theta})$, 
\begin{align}
\nabla f(y; \bar{\theta})&=\int q_{z|y}(z|y) \nabla \log{p_{z|y}(z|y; \bar{\theta})}dz=\int \nabla p_{z|y}(z|y; \bar{\theta})dz=0.  \label{deriv_f}
\end{align}
This yields that
\begin{align*}
E[\nabla \log p_b(\bar{\beta})\nabla^\top g(y; \bar{\theta})]&=E[\nabla \log{p_b}(\bar{\beta})\nabla^\top \log{p_y}(\bar{\theta})]=K_{b,y}. 
\end{align*}
Hence, by noting $g(y; \theta)=\log p_y(y; \theta)  + f(y; \theta)$, it holds that
\begin{align}
\mathcal{L}_x^{\rm cv}(\hat{\theta}_b)&=-\ell_{y}(\hat{\theta}_b)-\frac{1}{n}\sum_{i=1}^nf(y_i; \hat{\theta}_b)+\frac{1}{n}{\rm tr}(I_b^{-1} K_{b,y}) + o_p(n^{-1}). \label{cv_pf1}
\end{align}
For evaluating the second term on the right hand side, we apply Taylor expansion to $n^{-1}\sum_{i=1}^nf(y_i; \theta)$ around $\theta=\bar\theta$ 
by formally taking it as a function of $\beta$.
By noting (\ref{deriv_f}), this gives
\begin{align*}
\frac{1}{n}\sum_{i=1}^nf(y_i;\hat{\theta}_b)&=\frac{1}{n}\sum_{i=1}^nf(y_i; \bar{\theta})+ \frac{1}{2n}\sum_{i=1}^n(\hat{\beta}_b-\bar{\beta})^\top\nabla^2 f(y_i; \bar{\theta}) (\hat{\beta}_b-\bar{\beta}) + o_p(n^{-1})\\
&=\frac{1}{n}\sum_{i=1}^nf(y_i; \bar{\theta})+ \frac{1}{2n}{\rm tr}\left\{\sum_{i=1}^n\nabla^2 f(y_i; \bar{\theta}) (\hat{\beta}_b-\bar{\beta})(\hat{\beta}_b-\bar{\beta})^\top\right\} + o_p(n^{-1}). 
\end{align*}
It follows from the law of large numbers that
\begin{align*}
\frac{1}{n}\sum_{i=1}^n\nabla^2 f(y_i; \bar{\theta})&=\frac{1}{n}\sum_{i=1}^n\int q_{z|y}(z|y_i) \nabla^2 \log p_{z|y}(z|y_i; \bar{\theta}) dz \\
&\overset{p}{\rightarrow} E[\nabla^2 \log p_{z|y}(z|y; \bar{\theta})] = -I_{z|y}. 
\end{align*}
Hence, (\ref{theta_var}) indicates that 
\begin{align}
\frac{1}{n}\sum_{i=1}^nf(y_i;\hat{\theta}_b)&= \frac{1}{n}\sum_{i=1}^nf(y_i; \bar{\theta}) - \frac{1}{2n}{\rm tr}(I_{z|y}I_b^{-1}J_bI_b^{-1}) + o_p(n^{-1}). \label{cv_pf2}
\end{align}
By substituting (\ref{cv_pf2}) into (\ref{cv_pf1}), we establish that
\begin{align*}
\mathcal{L}_x^{\rm cv}(\hat{\theta}_b)&=-\ell_{y}(\hat{\theta}_b)+\frac{1}{n}{\rm tr}(I_b^{-1} K_{b,y}) + \frac{1}{2n}{\rm tr}(I_{z|y}I_b^{-1}J_bI_b^{-1}) -\frac{1}{n}\sum_{i=1}^n f(y_i; \bar{\theta}) + o_p(n^{-1}). 
\end{align*}
Hence, the proof completes. 
\end{proof}

\bibliographystyle{imsart-nameyear}
 \bibliography{IS19}

\end{document}